\documentclass[useAMS,usenatbib]{mn2e}
\usepackage{graphicx,amsmath,amssymb,subfigure,rotating,epsfig}
\usepackage{longtable}
\usepackage{pdflscape}
\usepackage{tabulary}
\usepackage{tabularx}

\bibpunct{(}{)}{;}{a}{,}{,}

\def\simgt{\mathrel{\lower0.6ex\hbox{$\buildrel {\textstyle >} \over {\scriptstyle \sim}$}}}
\def\simlt{\mathrel{\lower0.6ex\hbox{$\buildrel {\textstyle <} \over {\scriptstyle \sim}$}}}

\newcolumntype{C}[1]{>{\centering\let\newline\\\arraybackslash\hspace{0pt}}m{#1}}
\begin{document}

\title [MMAOs: An evolutionary sequence.]{6.7GHz Methanol Maser Associated Outflows: An evolutionary sequence.} 

\author[H. M. de Villiers et al.~2014] 
{H.\,M.\,de\,Villiers$^{1}$\thanks{Email: lientjiedv@gmail.com}, A.\,Chrysostomou$^{1}$, M.\,A.\,Thompson$^{1}$, J.\,S.\,Urquhart$^2$, S.\,L.\,Breen$^3$,\newauthor M.\,G.\,Burton$^4$, S.\,P.\,Ellingsen$^5$, G.\,A.\,Fuller$^6$, M.\,Pestalozzi$^7$, M.\,A.\,Voronkov$^8$,\newauthor D.\,Ward-Thompson$^9$ \\
\footnotesize 
$^1$Centre for Astrophysics Research, University of Hertfordshire, College Lane, Hatfield, Herts, AL10 9AB, United Kingdom\\
$^2$Max-Planck-Institut f\"ur Radioastronomie, Auf dem H\"ugel  69, D-53121 Bonn, Germany \\
$^3$CSIRO Astronomy and Space Science, Australia Telescope National Facility, PO Box 76, Epping, NSW 1710, Australia\\
$^4$School of Physics, University of New South Wales, Sydney, NSW 2052, Australia\\
$^5$School of Physical Science, University of Tasmania, Private Bag 37, Hobart 7001, TAS, Australia\\
$^6$Jodrell Bank Centre for Astrophysics, School of Physics and Astronomy, Alan Turing Building, \\~~University of Manchester,
Oxford Road, Manchester, M13 9PL, UK \\
$^7$IAPS - INAF, via del Fosso del Cavaliere 100, 00133 Roma, Italy \\
$^8$Australia Telescope National Facility, CSIRO Astronomy and Space Science, PO Box 76, Epping, NSW 1710, Australia \\
$^9$Jeremiah Horrocks Institute, University of Central Lancashire, Preston, Lancashire, PR1 2HE, United Kingdom \\
}

\maketitle
\begin{abstract}

We present a continuing study of a sample 44 molecular outflows, observed in $\rm{^{13}CO}$ lines, closely associated with 6.7GHz methanol masers, hence called Methanol Maser Associated Outflows (MMAOs). We compare MMAO properties with those of outflows from other surveys in the literature.   In general, MMAOs follow similar trends, but show a deficit in number at low masses and momenta, with a corresponding higher fraction at the high end of the distributions.  A similar trend is seen for the dynamical timescales of MMAOs.  We argue that the lack of relatively low mass and young flows in MMAOs is due to the inherent selection-bias in the sample, i.e. its direct association with 6.7GHz methanol masers.  This implies that methanol masers must switch on after the onset of outflows (hence accretion), and not before a sufficient abundance of methanol is liberated from icy dust mantles.  Consequently the average dynamical age of MMAOs is older than for the general population of molecular outflows.  We propose an adjusted evolutionary sequence of outflow and maser occurrence in the hot core phase, where methanol masers turn on \textit{after} the onset of the outflow phase.

\end{abstract}


\section{Introduction}
\label{sec:introduction}

The study of massive stars ($\rm{>8 M_{\odot}}$) is of importance, as they serve as principal sources of heavy elements, UV radiation, and feedback in the interstellar medium \citep[e.g.][]{Zinnecker2007}.  The formation and evolution process of young massive stars is complicated and not yet fully established.  In order to improve our understanding of the important events which take place during the formation of high-mass stars it is necessary to identify and study different evolutionary phases.

Our current understanding of the formation process of a massive star can be summarised into a crude four-stage evolutionary sequence \citep{Zinnecker2007}.  Prior to the optically visible main-sequence life of OB-type stars,  they spend $15\%$ of their lifetime in an embedded phase \citep{Churchwell2002}.  \citet{Zinnecker2007} sub-divide the embedded phase into four stages each of which is characterised by one or more types of objects.

The study presented here focuses on the Hot Molecular Core (HMC) stage, which is the first indirect manifestation of a young massive star prior to the well-studied UCH{\sc ii} phase.  HMCs are compact ($\leq 0.1$ pc), dense ($\geq 10^7 \rm{cm^{-3}}$), warm ($\geq 100$ K), and short-lived ($\leq 10^5$ yr) with high molecular brightness temperatures \citep[e.g.][]{Garay1999, Kurtz2000, Viti2005}. HMCs are warmed by an embedded (proto)star, but since the infalling material absorbs the UV flux from this central protostar \citep{Churchwell2002}, they show no or only weak radio free-free emission \citep{Beuther2005c,Dunham2011}.  

HMCs are also characterized by significant abundances of large, complex organic molecules, including methanol \citep[e.g.][]{vanDishoeck1998}.  During the initial cold ($\sim 10$ K) collapse phase, many molecular species that are present in the gas phase freeze out onto dust grain surfaces in the high-density inner region.  These grains provide a site for further chemical evolution, producing more complex species \citep[e.g.][]{vanDishoeck1998,Lintott2005}.  It is believed that the complex chemistry in HMCs comes from these icy mantles that build up on the dust grains during the early collapse phase \citep[e.g.][]{Brown1988,Nomura2004,Garrod2006}.  

Following this process, the associated protostar evolves to quickly warm the gas and dust (to $\sim 100$ K), evaporating the icy mantles, and injecting newly formed \citep[generally more complex;][]{Lintott2005} grain mantle material into the gas phase \citep[e.g.][]{vanDishoeck1998,Viti2004,Nomura2004,Garrod2006}.  According to \citet{Viti2004b}, this liberation of complex molecules into the gas phase is also assisted by molecular outflows developing from the central object, an important signpost of the HMC phase \citep[e.g.][]{Shepherd1996b,Cesaroni1997,Hunter1997,Kurtz2000}.  These outflows clear out cavities in the envelope, driving astrochemical processes via strong radiative (and possibly shock) heating of the cavity walls \citep[e.g.][]{vanDishoeck1998}. 

Hence, this is the stage when large amounts of methanol are released from the grain mantles \citep{Garrod2006}.  The period from the heating of low temperature grains to the desorption of ices is fairly brief,  $\sim 10^4 - 10^5$ yr \citep[e.g.][]{Bernasconi1996}.  \citet{Viti2004} found that organic molecules can be used as ``chemical clocks'' and showed that large species, such as methanol, are good indicators of the more evolved hot core phase, as these strongly bound species are abundant in the gas phase only at late times in a hot core's evolution.  They showed an increase in the fractional abundance of methanol by six orders of magnitude at $\sim 2 - 6 \times 10^4$ yr since warm-up started, for $10-25 M_{\odot}$ stars \citep[see also][]{Viti2005,Garrod2006,Garrod2008}.

Together with molecular outflows, interstellar 6.7GHz methanol masers are an important signpost of the HMC phase, being relatively common, the second brightest of the Galactic masers (only surpassed by 22GHz water masers), and radiating at cm wavelengths, hence not affected by high opacity gas and dust regions \citep[e.g.][]{Menten1991,Walsh1998,Minier2001,Minier2003,Beuther2002a,Walsh2003,Codella2004, Ellingsen2006,Purcell2006,Caswell2013}.  According to models by \citet{Sobolev1997b,Cragg2001,Sutton2001} and \citet{Cragg2005}, the radiative pumping of (class II) methanol masers by an infrared radiation source, happens in warm ($100~\rm{K} < T_{\rm{dust}}$), dusty environments, with gas densities between $10^6 - 10^9 \rm{cm^{-3}}$ and a high fractional methanol abundance of $> 10^{-7}$ (column densities of $\sim 10^{15} - 10^{17}~\rm{cm^{-2}}$) is required for 6.7GHz maser emission. HMCs have a brief lifetime, lasting between $2 \times 10^3$ and $6 \times 10^4$ years \citep{Kurtz2000}.  Work by \citet{Hartquist1995} and \citet{vanderTak2000} showed that it requires $10^4-10^5$ yr to obtain a high methanol abundance via evaporation of the icy dust grains in HMCs.  Given the required abundances for methanol masers and methanol abundance models, it is thus clear that methanol masers are likely to occur towards the end of the HMC phase.  Studying molecular outflows toward 6.7GHz methanol masers can reveal information regarding different sub-stages within the HMC phase, contributing to our developing understanding of the evolutionary timeline of massive protostars.

In \citet[][hereafter Paper I]{deVilliers2014}, high velocity wings were identified in 58 three dimensional position-position-velocity cubes, each simultaneously observing the $\rm{^{13}CO}$ and $\rm{C^{18}O}$ $\rm{(J = 3-2)}$ transitions with the HARP instrument on the JCMT, all targeted toward 6.7GHz methanol masers.  Usable maps existed for 55 of these spectra, and were utilised to calculate the physical properties of the outflows. Relations between outflows and their associated clump and maser properties were investigated and found to support the theory that massive star formation follows a scaled-up version of low-mass star formation.  Here, we compare the distributions of the physical properties of 44 Methanol Maser Associated Outflows (MMAOs) (see \S \ref{sec:observations}), with other CO outflow surveys.  Without redefining the whole evolutionary cycle for high mass star formation, the work presented here addresses the relationship between the 6.7GHz methanol masers, and the outflows they are associated with.

\section{Sample description and caveats}
\label{sec:observations}

An initial sample of 70 6.7GHz methanol maser positions, drawn from the Methanol Multibeam (MMB) catalogue \citep[][Breen et al., 2014 in prep.]{Green2009}, were observed with the heterodyne array receiver, HARP, on the James Clark Maxwell Telescope (JCMT) on the summit of Mauna Kea, Hawaii.  Targets were mapped in $\rm{^{13}CO}$ and $\rm{C^{18}O}$ $\rm{(J = 3-2)}$ transitions, covering a range of Galactic longitudes between $20.0^{\circ}$ and $34^{\circ}$. A total of 58 spectra and their maps, extracted at peak $\rm{^{13}CO}$ clumps, were of sufficient quality to be analysed.  The complete list of these target coordinates, and their 58 associated $\rm{^{13}CO}$ peak clump emission coordinates, is given in Paper I, together with a description of the data reduction and analysis.

All data cubes exhibited high velocity features, and $\rm{^{13}CO}$ integrated intensity maps were created using the blue- and red velocity wings.  Outflow maps were created for 55 targets (published in Appendix B of Paper I), and their physical properties tabulated (Table 5 of Paper I).  A subset of outflows were selected based on their having resolved kinematic distances and peak $\rm{^{13}CO}$ clump positions within $18''$ (c.f. the $14''$ JCMT beam at these frequencies) from their associated maser positions.  A total of 44 outflows satisfied these criteria, and were thenceforth referred to as Methanol Maser Associated Outflows (MMAOs). 

While it is not possible to resolve the contribution from individual stars or protostellar cores in our data, a relationship between the clump mass, the most massive star present in the clump, and the associated outflow's energetics can be assumed.   As discussed Paper I, the higher mass accretion rates of the most massive and dense clumps \citep{McKee2003} imply that they will arrive at the main-sequence ahead of their lower-mass neighbours \citep{Mottram2011,Urquhart2014b}.  Thus, while the most massive star, traced by the 6.7GHz methanol maser in this study, is close to joining the main sequence, it is likely that the lower-mass members of the proto-cluster still need to evolve to a stage where they can significantly contribute to the observed clump luminosity.  Results by \citet{Urquhart2013b} suggest that it is likely that the clump's bolometric luminosity is actually dominated by the most massive stars (see Paper I for more detail). \citet{Urquhart2014b} also found that there is a strong correlation between the clump masses and the bolometric luminosities of the most massive stars, but that the total clump luminosities are much lower than would be expected from the fully formed cluster.  This supports the hypothesis that the most massive stars have very rapid evolution times and are likely to dominate the observed clump luminosities, making it a reasonable assumption that so too, are the outflow luminosity and energetics.

\subsection{Outflow properties}
\label{sec:distribution}

We compare the distributions of the outflow masses $M_{\rm{out}}$, momenta $p$, and dynamical timescales $t_d$ with previous CO studies of massive molecular outflows, summarized in Table \ref{tab:litcatalogs}.  Although these samples are all biased in different manners, as described by their selection criteria, none of them are biased in terms of a $100 \%$ association with 6.7GHz methanol masers as is the case for MMAOs.  One could argue that the MMAO sample is the best defined, and most homogeneous of the four samples compared.  However, we take note that the methanol maser selection was carried out towards maser sources with accurate positions at the time, biasing them towards the more luminous sources, detected prior to the MMB surveys in targeted searches towards e.g. IRAS selected targets \citep[][and references therein]{Pestalozzi2005}.  This brightness bias will be discussed in more detail in \S \ref{sec:brightnessbias}.
\begin{figure}
		\begin{center}
		\includegraphics[width = 0.5\textwidth,clip]{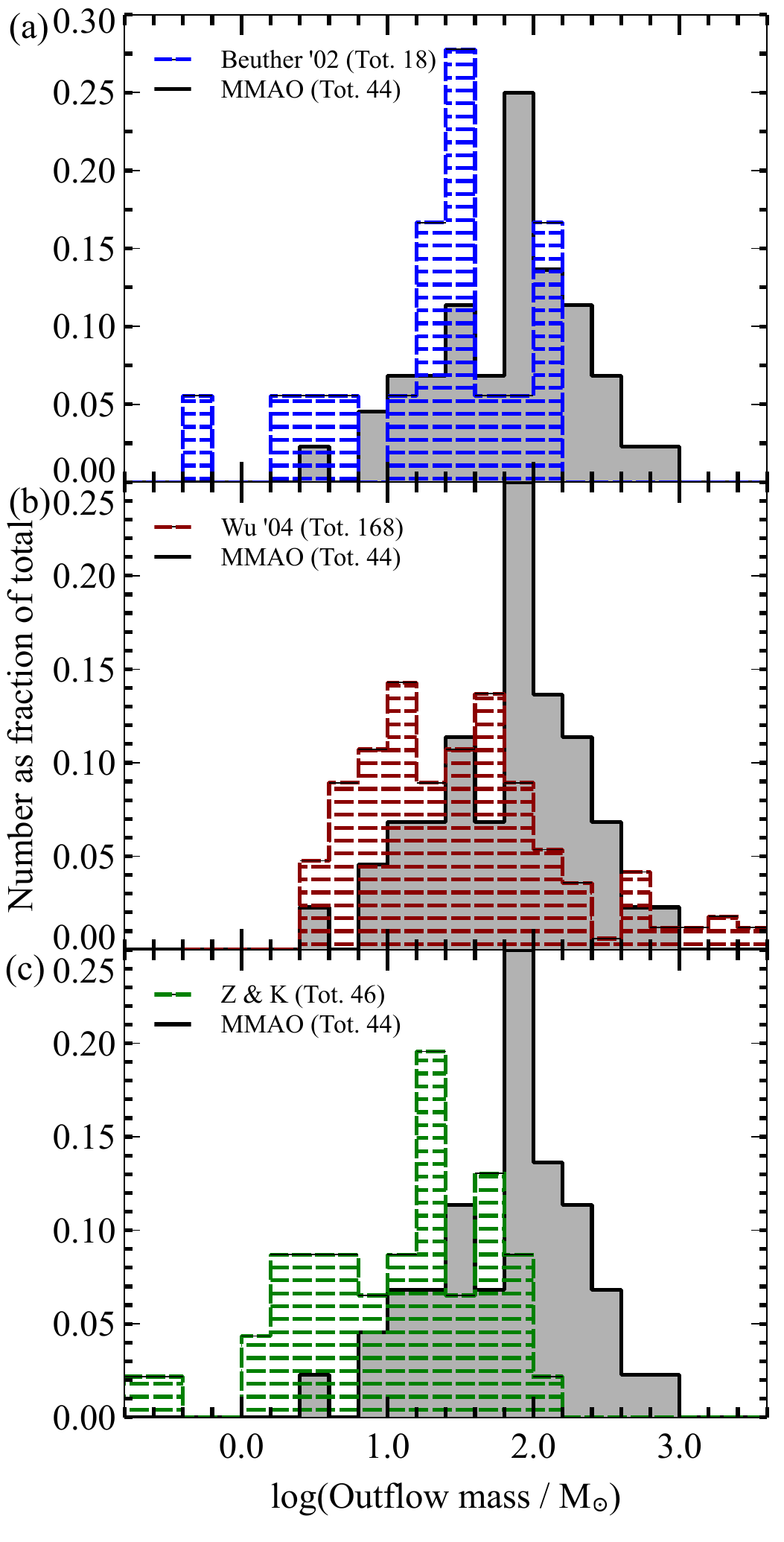} 
		\end{center}
		\caption{ \small{Comparison between the distributions of the logarithm of outflow masses for MMAOs (grey shading) and other surveys (horizontal dashed lines) (a) \citet{Beuther2002} (b) \citet{Wu2004} and (c) \citet{Zhang2005}.}}
		\label{fig:HistMass}
\end{figure}

\begin{figure}
		\begin{center}
		\includegraphics[width = 0.5\textwidth,clip]{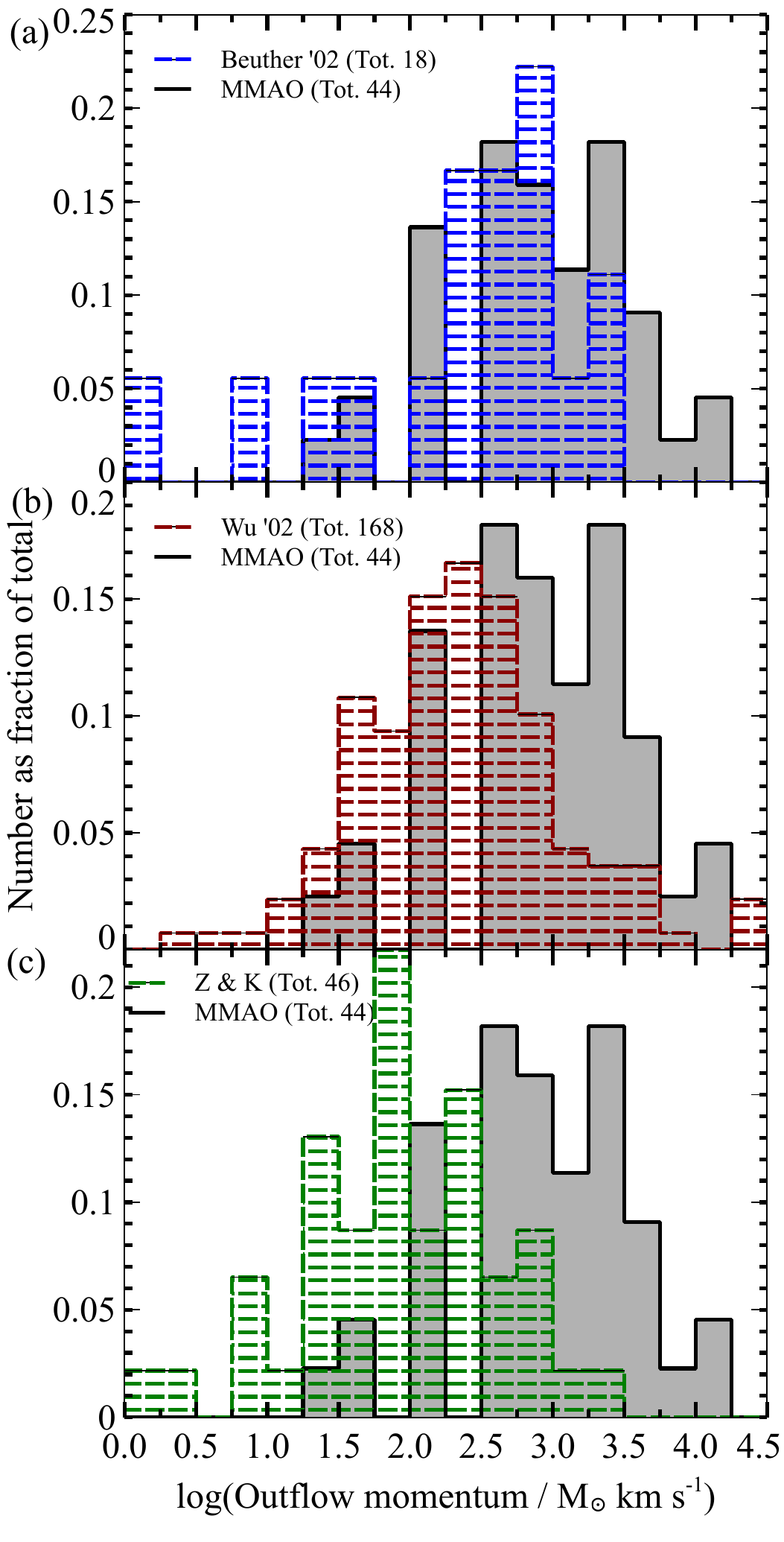}
		\end{center}
		\caption{ \small{Comparison between the distributions of the logarithm of the outflow momenta for MMAOs (grey shading) and other surveys (horizontal dashed lines) (a) \citet{Beuther2002} (b) \citet{Wu2004} and (c) \citet{Zhang2005}.}}
		\label{fig:HistMomentum}
\end{figure}

\begin{figure}
		\begin{center}
		\includegraphics[width = 0.5\textwidth,clip]{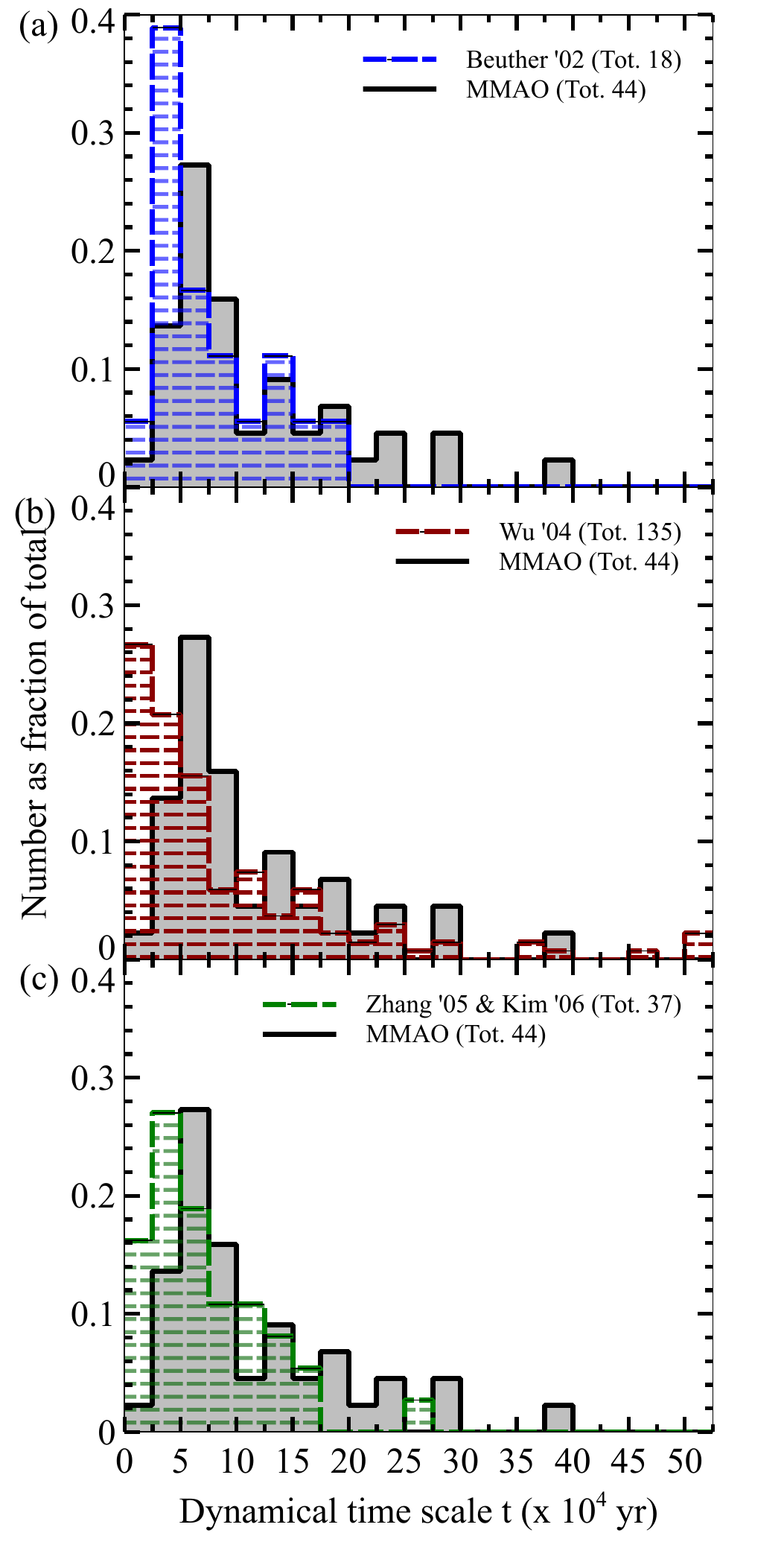} 
		\end{center}
		\caption{ \small{Comparison between the fractional abundance distributions of outflow dynamical timescales for MMAOs (solid lines) and other surveys (horizontal dashed lines) (a) \citet{Beuther2002} (b) \citet{Wu2004} and (c) \citet{Zhang2005}.}}
		\label{fig:HistTime}
\end{figure}

Figures \ref{fig:HistMass} to \ref{fig:HistTime} show the outflow mass, momentum and dynamical timescale distributions of MMAOs and the comparison data. The calculation of these parameters are discussed in detail in Paper I, and are briefly reviewed below.  For five\footnote{G 23.437-0.184; G 23.706-0.198; G 29.865-0.043; G 29.956-0.016 and G 31.282+0.062.} of the 6.7GHz methanol masers serving as tracers for MMAOs, trigonometric parallaxes have recently been published by \citet{Reid2014}.  These distances are updated from the values given in Paper I (calculated using the Galactic Rotation Curve), together with their corresponding outflow properties. These changes had an insignificant effect on the results presented in Paper I.

The outflow mass is calculated as 
\begin{equation}
	M_{\rm{out}} = \sum_{b,r} \left( N_i \times A_i \right)m_{\rm{H_2}},
	\label{eq:mass}
\end{equation}
where $b$ and $r$ represent the blue and red lobes respectively, $N_i$ the $\rm{H_2}$ column density of each lobe, calculated from the $\rm{^{13}CO}$ emission, with $T\rm{_{trans}}=31.8$ K, the upper level energy of the $\rm{J = 3-2}$ transition of $\rm{^{13}CO}$, and assuming an excitation temperature of $T\rm{_{ex}}=35$ K (see Paper I).  This $\rm{^{13}CO}$ column density is then converted to an $\rm{H_2}$ column density, assuming a Galactocentric dependant isotopic ratio \citep{Wilson1994}, and the $\rm{[CO]/[H_2]=10^{-4}}$ abundance ratio by \citet{Frerking1982}.  $A_i$ is the surface area of each lobe, and $m\rm{_{H_2}}$ is the mass of a hydrogen molecule.  

The momentum was calculated per velocity channel (width $\Delta v$) for each spatial pixel in the defined outflow lobe area, using the channel velocity relative to the systemic velocity ($v_i$), and the gas mass ($M_i$) corresponding to the emission in that channel.  This is summed over all velocity channels, and all pixels in each lobe area, giving:  
\begin{equation}
	p = \Delta v \left( \sum_{A_b} \left[ \sum_{i=v_b} M_{b_i} v_i \right] + \sum_{A_r} \left[ \sum_{i=v_r} M_{r_i} v_i \right] \right).
	\label{eq:momentum_me}
\end{equation}

The dynamical timescale $t_d$ is calculated as
\begin{equation} 
	t_d = \frac{l_{\rm{max}}}{\left( \Delta v_{b} + \Delta v_{r} \right) /2},
	\label{eq:tdyn}
\end{equation}
where $l_{\rm{max}}$ is the maximum lobe length between red ($l_r$) and blue ($l_b$) as measured from the clump coordinate to the furthest radial distance of the lobe. $\Delta v_b$ and $\Delta v_r$ are the velocity extents, measured from the peak velocity as defined by the $\rm{C^{18}O}$ spectrum, to the maximum velocity along the blue or red $\rm{^{13}CO}$ line wing (on a 3 $\sigma$ level above the noise).  In the case of monopolar outflow detections the length and velocity of the available lobe is used.  As described in Paper I, the values for $\Delta v_b$ and $\Delta v_r$ are scaled by a factor two, to account for the difference in velocity extent between $\rm{^{13}CO}$ (this work) and the wider $\rm{^{12}CO}$ profile used in the comparison studies.  For more detail regarding these measurements, we refer the reader to Paper I.

It is interesting to note that the mass distribution for MMAOs shows a relatively smaller fraction of low-mass outflows and a larger fraction of high-mass outflows compared to the other studies \citep[except for the few outliers from][]{Wu2004}.  The momentum distributions are shown in Figure \ref{fig:HistMomentum}, where the distribution for MMAOs have a similar fraction to the general outflow population for momenta between $\rm{10^2 M_{\odot}~km~s^{-1}}$ and $\rm{10^{2.75}~M_{\odot}~km~s^{-1}}$, but, as for outflow masses, low outflow momenta are under-represented. 

Figure \ref{fig:HistTime} shows the distribution for the dynamic timescale. Both \citet{Beuther2002} and \citet{Zhang2005} argue that for high-mass outflows, the dynamical timescale may often be a good estimate of the source age because the dynamical timescale of the outflow corresponds well to the free-fall timescale of the associated core.  The age range of MMAO outflows is consistent with the other work considered here, with the interesting difference being a lack of the youngest outflows at $\lesssim 5 \times 10^4$ yr.  

\onecolumn
\begin{landscape}
\begin{tiny}
\begin{table}
\centering
\caption{Other CO outflow studies used for comparison with the MMAO sample's masses, momenta and dynamical timescales.} \label{tab:litcatalogs} 

\begin{tabular}{|c|C{4cm}|C{3cm}|C{4cm}|C{3cm}|C{3cm}|} 
\hline
{} & \multicolumn{1}{c|}{\textbf{BEUTHER ET AL.}} & \multicolumn{1}{c|}{\textbf{WU ET AL.}} & \multicolumn{1}{c|}{\textbf{ZHANG ET AL.}} & \multicolumn{1}{c|}{\textbf{KIM \& KURTZ}} & \multicolumn{1}{c|}{\textbf{MMAOs}} \\\hline 

\textbf{Outflow sample size} & Used 18 out of 24, which had resolved distances. & 168 for $M_{\rm{out}}$; \newline 139 for $p$; 135 for $t_d$ & 35 & 11 & 44 \\
\hline
\textbf{Molecular tracer} & $\rm{^{12}CO(J=2-1)}$ & $\rm{^{12}CO(J=1-0)}$ and $\rm{^{12}CO(J=2-1)}$ & $\rm{^{12}CO(J=2-1)}$ &  $\rm{^{12}CO(J=2-1)}$ & $\rm{^{13}CO(J=3-2)}$ \\
\hline
\textbf{Facility} & 30 m IRAM \newline at Pico Veleta & Literature study & 12 m NRAO at Kitt Peak \& \newline 10.4 m CSO & 12 m NRAO at Kitt Peak & 15m JCMT at Mauna Kea \\
\hline
\textbf{Resolution} & Vel: 0.1 $\rm{km~s^{-1}}$ smoothed to 1 $\rm{km~s^{-1}}$ \newline Spatial: $11''$ \newline (0.07 - 0.56 pc) & Mixed & Vel(NRAO): \newline 1.02 $\rm{km~s^{-1}}$ (run 1) \& \newline 0.5 $\rm{km~s^{-1}}$ (run 2) \newline Spatial(NRAO): $29''$ \newline Spatial(CSO): $30''$ \newline (0.04 - 1.57 pc) & Vel: 0.68 $\rm{km~s^{-1}}$ \newline Spatial: $27''$ \newline (0.06 - 0.60 pc) & Vel: 0.5 $\rm{km~s^{-1}}$ \newline Spatial: $6''$ \\
\hline
\textbf{Sensitivity} & Int.time $\leq 5$ min.; rms: Not given. & Mixed & NRAO: Int. time = 4 min (run 1) \newline \& 1 min (run 2); \newline rms = 0.2 K/2 $\rm{km~s^{-1}}$ \newline CSO: rms = 0.03 K/1.5 $\rm{km~s^{-1}}$ & rms = 0.2 K & rms = 0.24 K\\
\hline
\textbf{Selection criteria} & A sample of high mass protostellar objects from the \textit{IRAS} catalog, following \citet{Wood1989} criteria, refined by \citet{Ramesh1997} for the identification of UCH{\sc ii} regions. & Combined study of 391 high velocity molecular outflows from literature published prior to February 2003. High-mass outflows satisfies $ L_{\rm{out}} > 10^3 \rm{L_{\odot}}$ and $M_{\rm{out}}\geq 3  \rm{M_{\odot}}$. & A selection from the flux-limited sample of bright \textit{IRAS} sources, originally 260, following requirements from \citet{Palla1991}, 101 towards which \citet{Molinari1996} found $\rm{NH_3}$ emission, and \citet{Molinari1998} observed 67 of them for radio continuum with VLA. \citet{Zhang2005} used these 67, with additional two in their search. & \citet{Molinari1998} did not study 34 of the sources from \citet{Molinari1996}, of which Kim \& Kurtz selected 11 and added one additional source. & Outflows associated with 6.7GHz methanol masers, where the latter have resolved kinematic distances and are located within $18''$ of the peak $\rm{^{13}CO}$ emission.\\
\hline
\textbf{References} & \citet{Beuther2002,Sridharan2002} & \multicolumn{1}{c|}{\citet{Wu2004}} & \multicolumn{1}{c|}{\citet{Zhang2005}} & \multicolumn{1}{c|}{\citet{Kim2006}} & \multicolumn{1}{c|}{\citet{deVilliers2014}}\\
\hline
\textbf{Notes} & The ambiguous distance to three sources was resolved using \citet{Roman2009}. & & & & \\

\hline
\end{tabular}
\end{table}
\end{tiny}
\end{landscape}
\twocolumn

\section{Discussion} 
\label{sec:discussion}

Considering Figures \ref{fig:HistMass} to \ref{fig:HistTime}, there seems to be an under-abundance of MMAOs with low mass, momentum and dynamical timscales.  The latter is of particular interest.

\begin{figure}
		\begin{center}
		\includegraphics[width = 0.4\textwidth,clip]{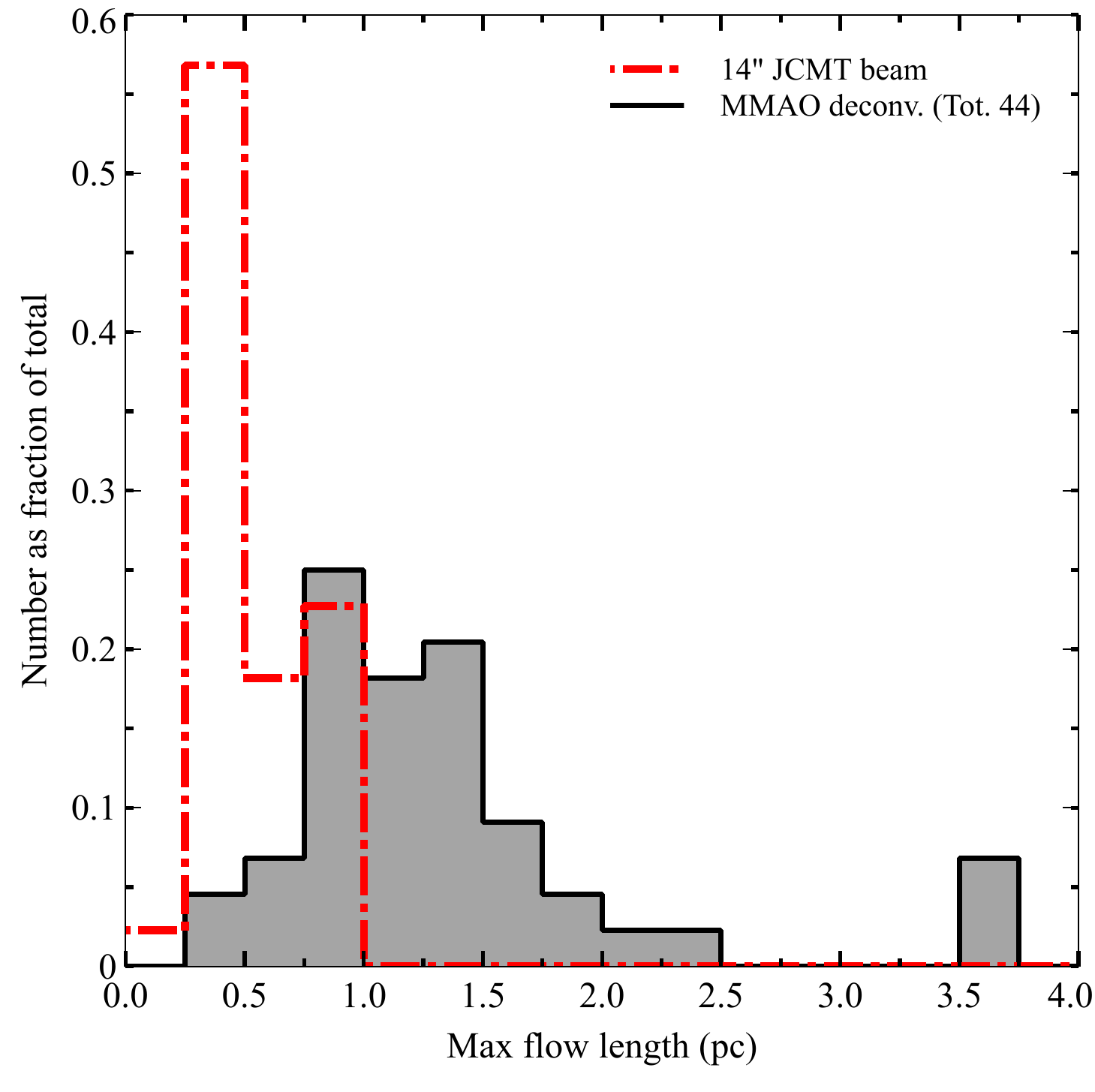} \\
		\end{center}
		\caption{ \small{Distributions of the maximum outflow length for each MMAO (shaded area) and the physical lengths for the resolution limit of a $14''$ JCMT beam given the distance to each MMAO target (red thick dashed line).}}
		\label{fig:HistLengthbeam}
\end{figure}

The two main independent variables in the calculations of outflow mass and dynamical timescale, are the outflow lobe's projected size (included via the outflow's area $A_{\rm{r/b}}$ in the mass), or $l_{\rm{max}}$ in the dynamical timescale, and the average velocity extent $(\Delta v_{\rm{avg}} = \Delta v_{b} + \Delta v_{r}) /2$.  An important question to answer is whether the comparatively high lower limit and high mean of MMAOs' outflow lengths, is a resolution effect, or does it reflect an actual property of the sample? The same question can be asked about the lack of low mass and young outflows in our MMAO sample. Since the study has a $100\%$ outflow detection rate (Paper I), these effects cannot be due to non-detections.  However, since the full width half maximum (FWHM) of the JCMT beam is $\sim 14''$, outflow lobes with an angular size smaller than this will not be resolved, imposing a limit on the smallest outflow sizes measured.  In order to determine whether the observed distribution is affected by the telescope's resolution, we (a) deconvolved the outflow lobe diameters with a $14''$ beam and re-measured the lobe lengths (Paper I), and (b)  used the angular resolution limit of the JCMT as a hypothetical physical lobe size at each MMAO's distance.  The distributions of the maximum deconvolved lobe length for each MMAO outflow (shaded area) and the linear size of a JCMT beam at each MMAO distance (red dashed line), are shown in Figure \ref{fig:HistLengthbeam}. The fact that the MMAO distribution shows a deficit in short outflow lengths, beyond the distance limit defined by the telescope's resolution, indicates that such a deficit is not instrumental in nature, but is real.  That is, if smaller outflows did exist in this sample of MMAOs, they would have been detected.

\begin{figure}
		\begin{center}
		\includegraphics[width = 0.4\textwidth,clip]{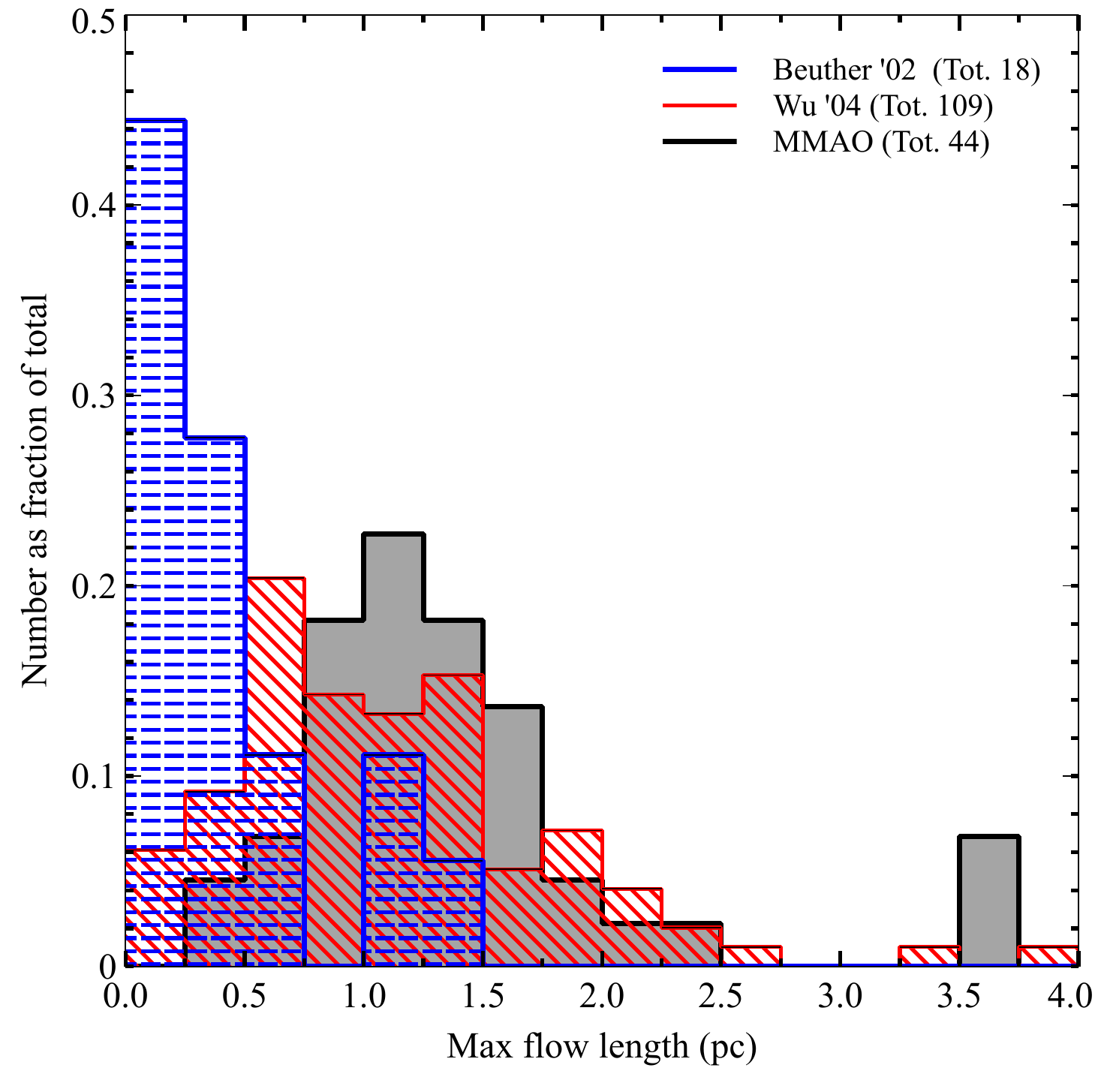} 
		\end{center}
		\caption{ \small{Distributions of the maximum outflow length for MMAOs (shaded area), \citet[][blue horizontal lines]{Beuther2002}, and \citet[][red diagonal lines]{Wu2004}.}}
		\label{fig:HistLength}
\end{figure}

The maximum lobe length ranges from 0.3 pc to 3.7 pc for MMAOs, with a mean of 1.2 pc (see Table 3 in Paper I).  These lengths are of a similar order of magnitude to both the samples of \citet{Beuther2002} and \citet{Wu2004}, although they cover a larger range and have a higher mean than the outflow lengths from \citet{Beuther2002}, which cover a range of $0.1-1.8$ pc, with a mean of 0.8 pc.  The MMAO values compare well with those from \citet{Wu2004}, ranging from $0.04 - 3.98$ pc, with a mean of 1.1 pc.  The distributions of these lengths are shown in Figure \ref{fig:HistLength}, which does not show the outflow lobe lengths for the \citet{Zhang2005} and \citet{Kim2006} samples, as no lengths are published by \citet{Zhang2005}, leaving only the eleven outflow lengths published by \citet{Kim2006} for that combined sample, which would not be statistically significant. \citet{Beuther2002} detected a higher fraction of small outflow sizes compared to the MMAOs, which potentially contributes to the bottom-heavy distributions of their outflow masses and dynamical timescales.

The average values and distribution of the outflow velocities, $\Delta v_{\rm{avg}}$ are found to range between $5 - 31~\rm{km~s^{-1}}$, with a mean of $13~\rm{km~s^{-1}}$ for MMAOs.  \citet{Beuther2002} has a similar range and mean, from 4 to $27~\rm{km~s^{-1}}$ with a mean of $13~\rm{km~s^{-1}}$.  In \citet{Wu2004}, a wider range of $3 - 133~\rm{km~s^{-1}}$ is seen, but again a similar mean of $18~\rm{km~s^{-1}}$ and median of $13~\rm{km~s^{-1}}$ (we calculated the latter as the Wu et al. sample is skewed by a few high velocity outliers). Note that these values depend on the sensitivity of the observations to outflow wings, but as mentioned in \S \ref{sec:distribution}, our $\rm{^{13}CO}$ measurements are scaled to equivalent values expected in $\rm{^{12}CO}$ (see Paper I). The velocity distributions were found to be similar for all samples, and the fact that the outflow mass and momentum distributions are similarly top-heavy for MMAOs, implies that the velocity does not have much of an effect towards changing the shape of the MMAOs' dynamical timescale distribution. No data is available for \citet{Zhang2005} and \citet{Kim2006}.

\begin{figure}
		\begin{center}
		\includegraphics[width = 0.4\textwidth,clip]{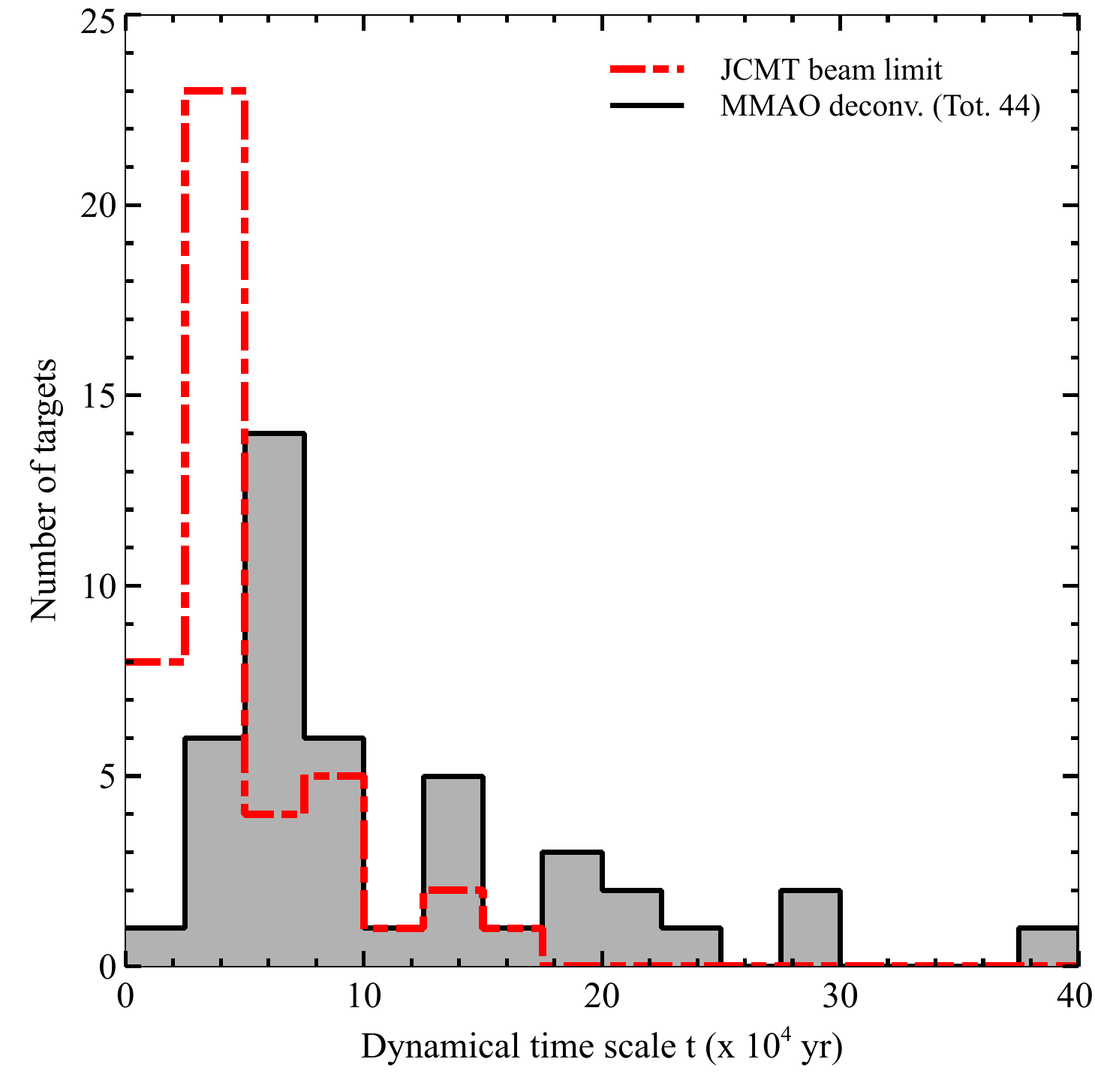} \\
		\end{center}
		\caption{ \small{Distributions of the dynamical timescales for MMAOs after beam deconvolution (shaded area), and if every target only had a minimum lobe size of the $14''$ JCMT beam (red dashed line).}}
		\label{fig:HistTime_beam}
\end{figure}

Figure \ref{fig:HistTime_beam} shows the distribution of dynamical timescales for our MMAO sample after the data have been deconvolved (the latter had minimal impact on the distribution), as well as the dynamical timescales calculated for minimum lobe length the size of the JCMT beam, corresponding to Figure \ref{fig:HistLengthbeam}.  It shows that our MMAO sample, as observed from the JCMT, is not sensitive (i.e. is incomplete) for ages below $\sim 2.5 \times 10^4$ yr. MMAOs show a turn-over at $\sim 5 \times 10^4$ yr, implying that the lack of the youngest outflows is not due to a resolution limitation.

\begin{figure}
		\begin{center}
		\includegraphics[width = 0.4\textwidth,clip]{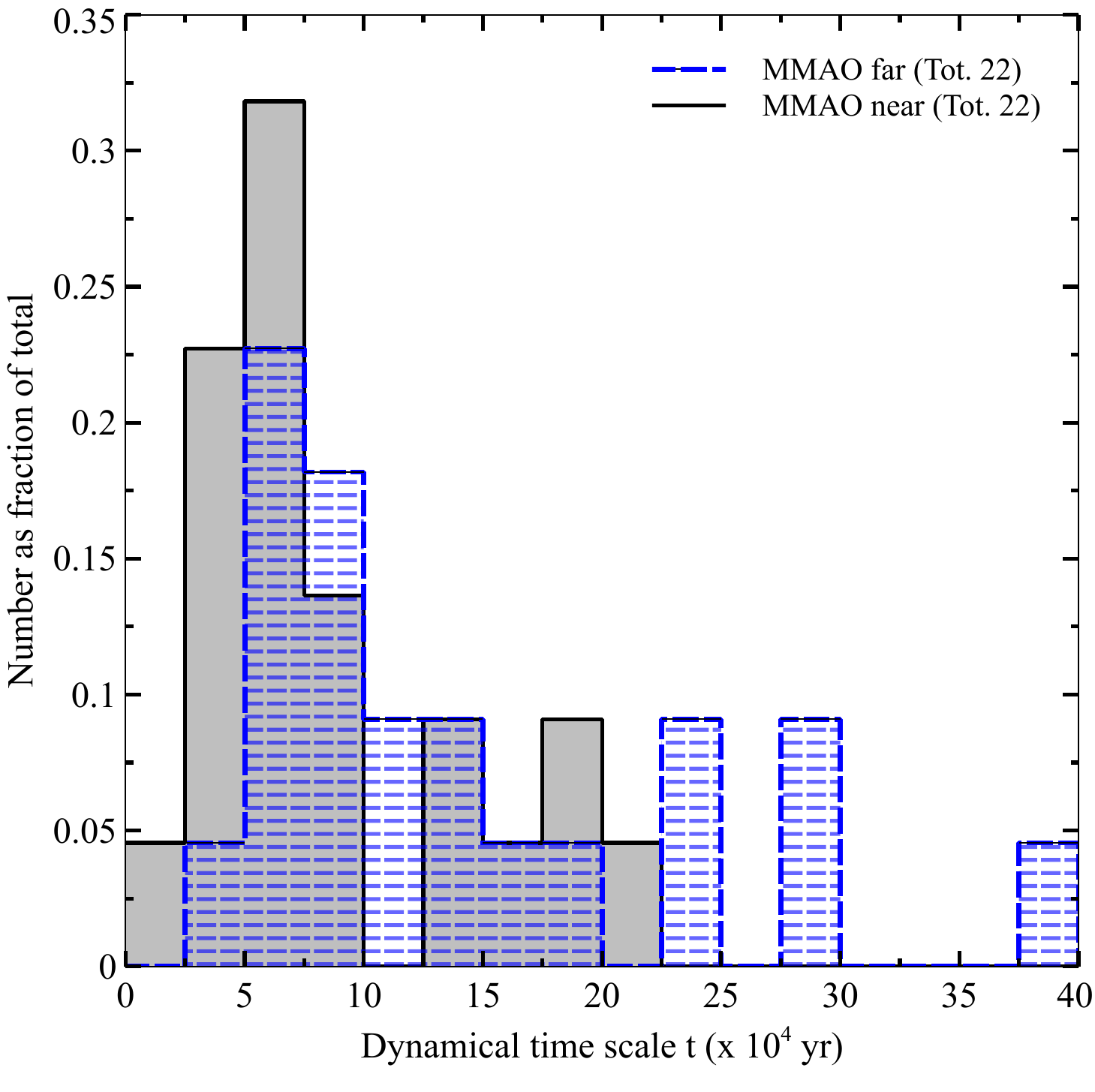} \\
		\end{center}
		\caption{ \small{Jackknife test for a Malmquist bias in the dynamical timescale distribution of MMAOs.}}
		\label{fig:HistTime_jackknife}
\end{figure}

We also investigated whether a distance bias could cause the observed turn-over at low dynamical timescales.  This is a risk because one tends to observe only the brightest sources at the greatest distances (Malmquist bias).  To investigate whether our data is affected by such a bias, we performed a jackknife test, ranking our dataset according to distance, and dividing it in half to form two distributions (Figure \ref{fig:HistTime_jackknife}).  Both the ``near'' and ``far'' samples show a turn-over in dynamical timescales at $\leq 5 \times 10^4$ yr, indicating that it is unlikely to be a distance-dependent effect.

Finally, we consider the inherent selection bias in our data, as \textit{all} our targets are associated with 6.7GHz methanol masers.  If the differences in the outflow mass and timescale distributions for MMAOs are due to this selection bias, one might argue that this makes our sample's properties distinct from those of others.  

Although not a simple process, we expect that for any individual outflow, mass will increase with time as long as accretion continues to power the expanding flow.  The fractional distribution of MMAO mass favours large values, compared to the other samples (Figure \ref{fig:HistMass}), supporting the suggestion that the average age of MMAOs is older than a sample not biased toward 6.7GHz methanol maser emission.

The methanol maser association of our sample implies the presence of complex organic molecules in the gas phase at high column density. \citet{Viti2004} found that large organic molecules such as methanol are good indicators of the more evolved hot core phase.  Both \citet{Hartquist1995} and \citet{Viti2004} state that methanol shows a rapid increase to a high enough fractional abundance for the maser to switch on around $10^4-10^5$ years after infall and consequent warm-up commenced.  This implies that there is a time in the hot core evolution, \textit{before} which masers do not exist due to an insufficient fractional abundance of methanol.  The turn-over seen in the MMAO age distribution in Figure \ref{fig:HistTime} is consistent with these ages.  

The evolutionary sequence of masers as given by \citet[][their Figure 6]{Breen2010} suggests that OH masers usually trace a more evolved stage of star formation compared to other masers \citep[e.g.][]{Garay1999}.  We used the catalogs in \citet{Caswell2013} and \citet{Qiao2014} to match our MMAOs with OH masers, and found an association within $\sim 3.5''$ for at least $50 \%$ of the MMAOs.  This is consistent with the suggestion that they occur at a later time in the hot core phase, although, given \citet{Breen2010}'s diagram, some 6.7GHz masers are also expected to occur earlier than the OH maser stage, which could account for the other $50 \%$ of our sample.

Hence, we conclude that this study's association with methanol masers introduces an age-bias towards more evolved outflows, suggesting that masers switch on sometime after the outflow has started.  This implies that outflows found to be associated with 6.7GHz masers, should generally be slightly more evolved than massive outflows not associated with these masers, and that it is unlikely that any of the very youngest (and probably also the smallest) outflows would appear in our sample, given its bias.  

\subsection{Maser brightness bias}
\label{sec:brightnessbias}
The area of the sky from which our sample of MMAOs happen to be selected includes a region of very high star formation activity in the Galaxy, (i.e. mini-starburst W43 and the intersection of the end of the bar and one of the spiral arms).  It is important to understand whether the sample could have been biased towards higher maser luminosities by this, and if so, to what extent this could influence the results.  We found that the maser peak luminosities for MMAOs (Breen et al. in prep.), cover a range of $\sim 2.5$ orders of magnitude, while the MMB 6.7GHz masers cover a luminosity range up to $\sim 6$ orders of magnitude \citep[][in their Figure 7]{Breen2011}.  

\begin{figure}
		\begin{center}
		\includegraphics[width = 0.45\textwidth,clip]{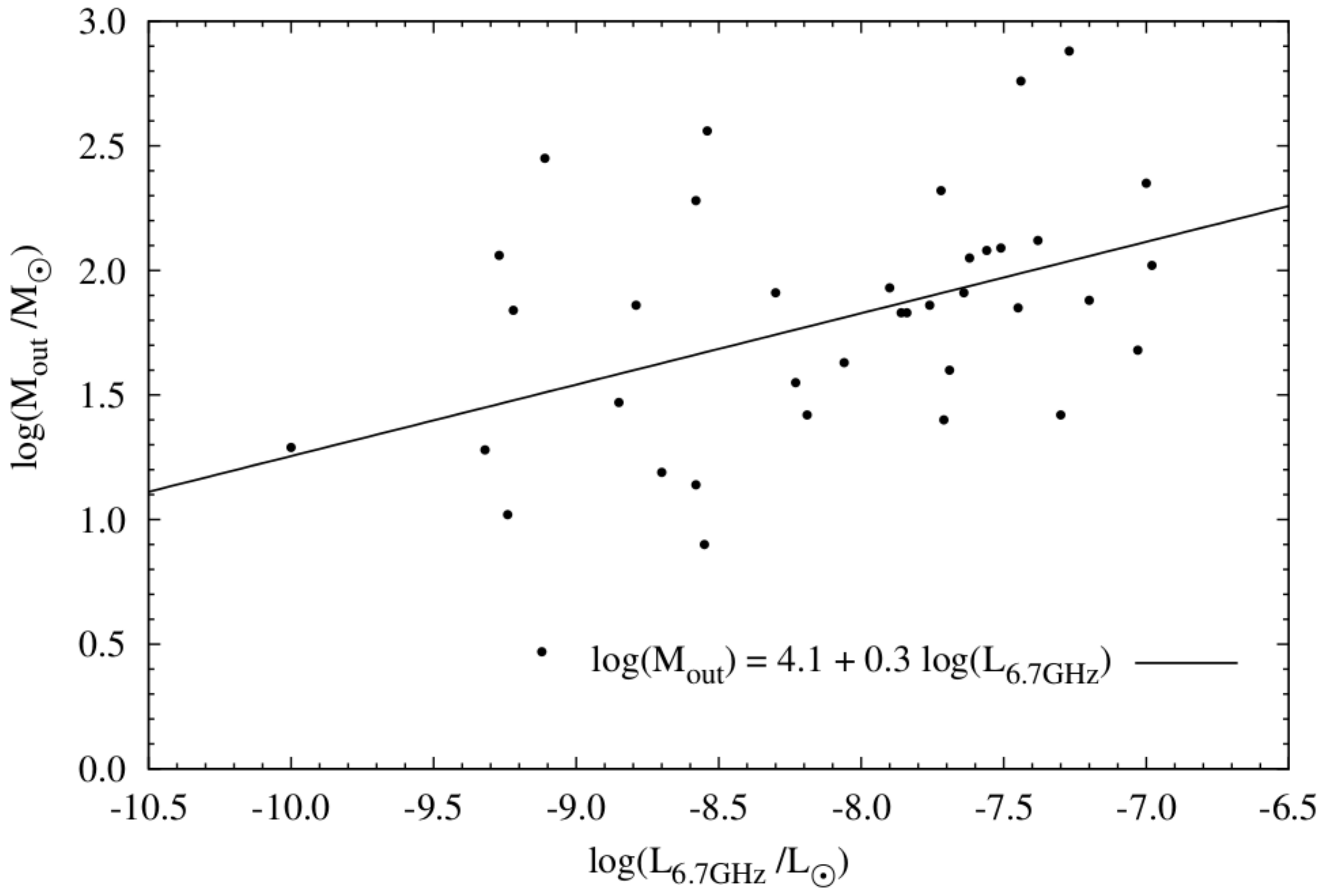} 
		\end{center}
		\caption{ \small{Outflow mass as a function of the associated 6.7GHz maser's luminosity for all MMAOs with available 6.7GHz luminosity data (Breen et al. in prep.).}}
		\label{fig:Mout_vs_Lmas}
\end{figure}

\begin{figure}
		\begin{center}
		\includegraphics[width = 0.45\textwidth,clip]{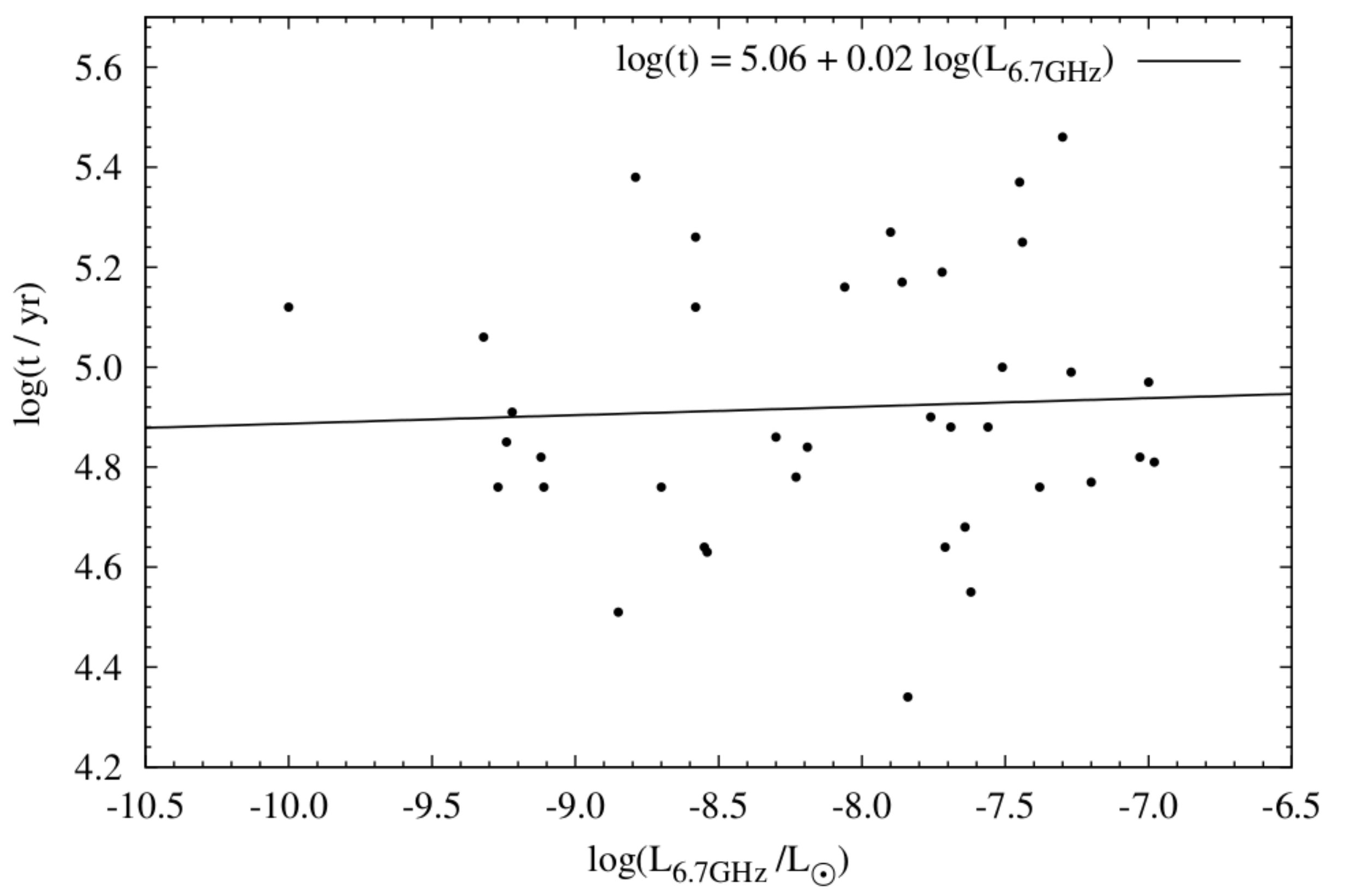} 
		\end{center}
		\caption{ \small{Outflow dynamical timescales as a function of the associated 6.7GHz maser's luminosity for all MMAOs with available 6.7GHz luminosity data.}}
		\label{fig:t_vs_Lmas}
\end{figure}

Although we do not sample below a 6.7GHz maser luminosity of $100~\rm{Jy~kpc^2}$ \citep[calculated as in][their Figure 14]{Urquhart2013}, and an investigation of the $L_{\rm{maser}}/M_{\rm{clump}}$ distribution shows a bias to higher 6.7GHz maser luminosities for the MMAOs, our sample spans the luminosity range in which the bulk of the ATLASGAL population lies.  

We investigate whether a correlation exists between the maser luminosity and outflow mass and dynamical timescale in Figures \ref{fig:Mout_vs_Lmas} and \ref{fig:t_vs_Lmas}. Only a weak power-law relation exists for the outflow mass, $M_{\rm{out}} = (1.3 \times 10^4)L_{\rm{6.7GHz}}^{0.3}$, with a Spearman Rank coefficient of $r_s = 0.44$ ($t_s = 2.96$, where the minimum value of $t_s$ for the correlation to be significant is $t_{s_{\rm{crit}}} = 2.02$ -- see Paper I for more detail).  It is thus possible that the MMAO sample may have a slight bias towards more massive outflows, albeit with a weak correlation and substantial scatter in the data.

Within our sample, no significant correlation is found between $t_{\rm{dyn}}$ and $L_{\rm{6.7GHz}}$, with a power-law fit of $t_{\rm{dyn}} = (1.15 \times 10^5)L_{\rm{6.7GHz}}^{0.02}$ ($r_s = 0.08$ and $t_s = 0.45$), although we note that our sample only covers the brighter end of the MMB luminosity function. \citet{Breen2010} proposed that the 6.7GHz maser luminosity increases as the source evolves.  On the other hand, \citet{Urquhart2013} suggested that the strength of the maser emission is dominated by the energy output of the central source and not driven by source evolution.  Either way, any maser luminosity bias should be normalized out in $t_{\rm{dyn}}$ calculations, as is seen in Figure \ref{fig:t_vs_Lmas}, since for a given age, the more energetic sources will have larger sizes ($l_{\rm{max}}$), but their outflow velocities ($v_{\rm{max}}$) will also be faster. This implies that the lack of ``younger'' outflows in Figure \ref{fig:HistTime} should persist regardless of a maser luminosity bias. 

\subsection{A revision to the evolutionary scheme for the hot core phase}
\label{sec:timeline}

\citet{Codella2004} undertook a $\rm{^{13}CO}$ survey, searching for molecular outflows toward UCH{\sc ii} regions, for which they already had 6.7GHz methanol and 22.2 GHz water maser data available.  Based on the results, they proposed a tentative evolutionary scheme for the formation process of massive stars that suggests maser emission occurs before the outflow is developed enough to be detected.  Following this, both the maser and outflow is present and detectable, after which maser emission disappears while the outflow is still present.  Finally, only the UCH{\sc ii} region without masers or outflows is present. This sequence can be divided in the following steps:
\begin{enumerate}
	\item [(1)] the earliest phase: maser emission present, outflow not yet developed enough to be detected,
	\item [(2)] both the maser and outflow are present and detectable,
	\item [(3)] maser emission disappears while the outflow remains,
	\item [(4)] outflow switches off, only the UCH{\sc ii} region remains.
\end{enumerate}
They argued that the lack of outflows in phase (1) does not imply a real lack of outflows, but instead reflects the youthfulness of the source whose outflow is too small to be detected by single-dish observations.  Contrary to this stage where \citet{Codella2004} do not detect an outflow associated with every 6.7GHz methanol maser, we found a $100 \%$ outflow detection rate towards 6.7GHz methanol masers in Paper I, which could be due to the higher sensitivity of our survey (the average rms of their $\rm{^{13}CO~J=2-1}$ spectra is 0.5 K versus our 0.24 K -- see Paper I).  \citet{Codella2004} also showed that, if the outflow ages obtained in various surveys \citep{Shepherd1996b,Shepherd1996a,Beuther2002}, lying between $10^4 - 10^5$ yr, are assumed to be representative estimates of their observed sample, and if the expected lifetime of an UCH{\sc ii} region is $\sim 10^5$ yr \citep{Wood1989}, their time-line implies the maser phase should end after a few $10^4$ yr, as the masers are expected to turn off once the expanding UCH{\sc ii} region dissipated the methanol below critical masing levels, \citep[e.g.][]{Walsh1998,Walsh2003}.

From our results presented in \S\ref{sec:distribution}, supported by \citet{Viti2004}'s chemical evolution models, we propose an amendment to this scheme, in which the maser switches on after the outflow has started and off before it disappears:
\begin{enumerate}
	\item [(1)] the earliest phase: outflow develops and grows with no methanol maser present; the infall process heats the dust grains and releases methanol into the gas phase, 
	\item [(2)] both methanol maser and outflow are present and detectable (sufficient column density for methanol maser to turn on), 
	\item [(3)] methanol maser emission disappears while the outflow remains, 
	\item [(4)] outflow switches off, only the UCH{\sc ii} region remains.
\end{enumerate}
Thus, we propose that 6.7GHz methanol masers seem to signpost a more advanced stage of the hot core phase, and not as early as sometimes argued \citep[e.g.][]{Codella2004}.  Note that we only propose a change to the beginning of the evolutionary scheme, not the end, which we have assumed stays unchanged.

\subsection{Other considerations}

Although MMAOs represent a homogeneous sample of objects (i.e. observed with the same telescope and instrument), as with almost any statistical sample, there will be outliers and exceptions.  Furthermore, these stages in the evolutionary sequence of a massive YSO are not always clear-cut, and it is very possible for different evolutionary stages to overlap \citep[e.g.][]{Beuther2007}. One example is G 24.790+0.083 which is in the MMAO sample, but shows associations with Class I and II methanol, water and OH masers \citep{Moscadelli2007}.  OH masers are usually indicative of a more evolved stage in the evolution of a massive YSO \citep{Caswell1997}, and yet this source has been calculated to have a relatively ``young'' age of $\sim 2 \times 10^4$ yr \citep{Beltran2011}, younger than those calculated in Paper I.  Similarly, G 23.010-0.411 has an age of $1.8 \times 10^4$ yr \citep{Sanna2014}, and is also associated with Class I and II methanol, water and OH masers \citep{Forster1999,Cyganowski2009}.

\begin{figure}
		\begin{center}
		\includegraphics[width = 0.4\textwidth,clip]{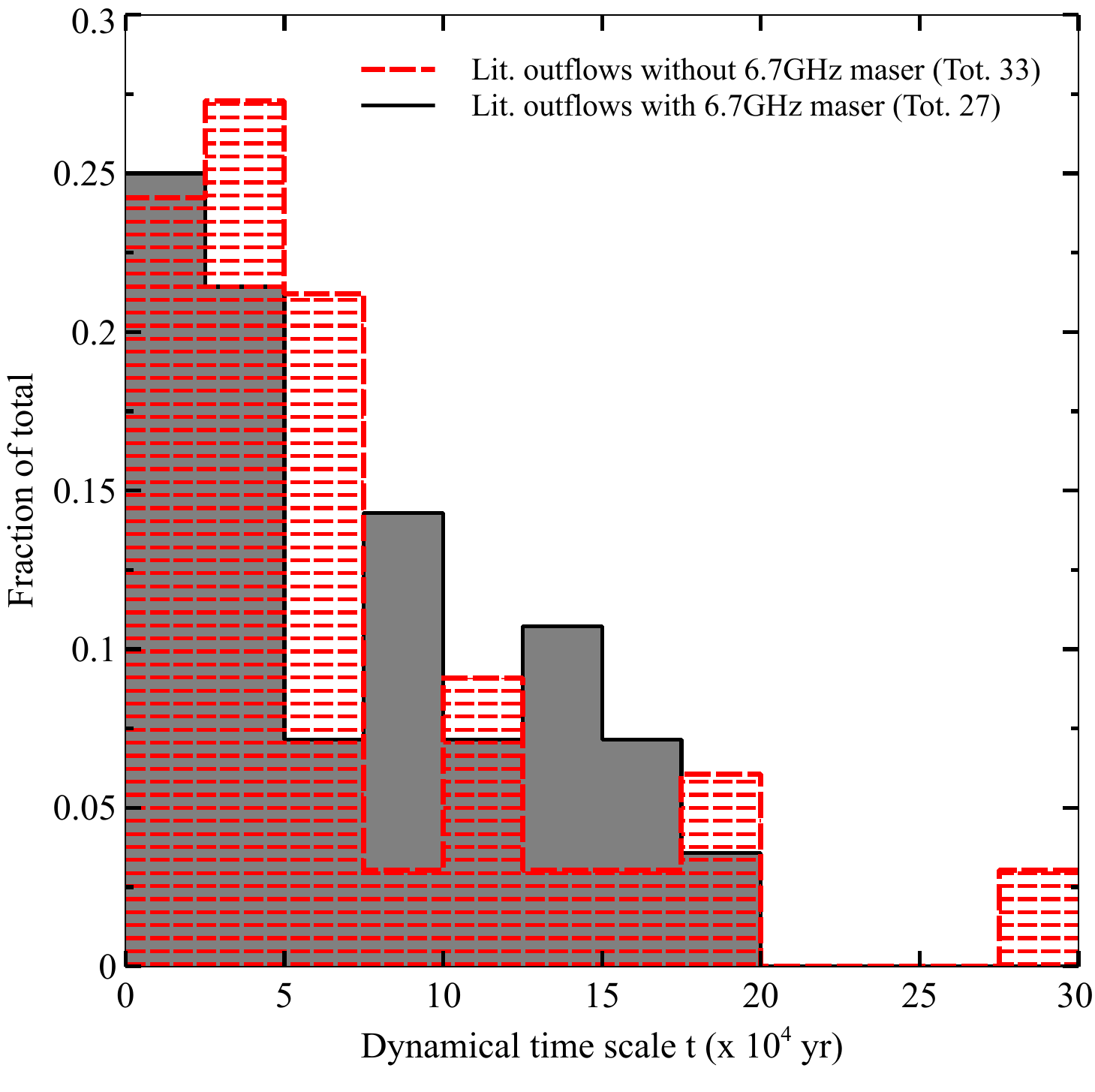} 
		\end{center}
		\caption{ \small{Distributions of the dynamical timescales of the outflows from the surveys used for comparison in this study for whom 6.7GHz methanol maser association data were available.  The grey shaded area shows the outflows \textit{with} methanol masers associated, and the red dashed lines show the outflows \textit{without} methanol masers.}}
		\label{fig:Hist_timesmasers}
\end{figure}

We investigated whether a similar trend to that of the MMAOs in Figure \ref{fig:HistTime} exists for methanol maser associated outflows in the literature.  Outflows from \citet{Wu2004} were cross-matched with the Methanol Multibeam Survey \citep[MMB,][]{Green2009}, and The Arecibo Methanol Maser Galactic Plane survey \citep[AMGPS,][]{Pandian2007} catalogues, with matching radii the size of the JCMT beam.  Unfortunately, these two catalogues contain maser positions that were observed in the Southern Hemisphere, while most outflow surveys were observed in the North, implying a small cross-over area.  Where information regarding 6.7GHz methanol maser association were available for the \citet{Beuther2002,Zhang2005} and \citet{Kim2006} samples, we added them to the above two datasets.  We found 27 outflows associated with masers, and 33 without.

The distribution of these outflows' dynamical timescales is shown in Figure \ref{fig:Hist_timesmasers}. Compared
to outflows without associated methanol masers, no deficiency in outflows \textit{with} methanol masers is seen at young ages ($\leq 2.5 \times 10^4$ yr), as is found for the MMAOs.  However, a KS-test has shown that the null hypothesis that these two datasets are drawn from the same sample, cannot be rejected, and the difference seen between them is consequently statistically insignificant.  We recognise that Figure \ref{fig:Hist_timesmasers} shows the presence of outflows associated with masers at young ages, whereas they appear to be absent for MMAOs at similar ages. However, we have shown that should these young outflows be present in the MMAO sample, they would have been detected (Figure \ref{fig:HistTime_beam}), yet they are not.  Furthermore, the samples plotted in Figure \ref{fig:Hist_timesmasers} are very heterogeneous (different observation techniques and calculation methods), while the advantage of the MMAO study is that observations and analyses were carried out in a homogeneous manner. 

In the near future, we should be able to assess these issues with stronger statistical support by expanding the MMAO sample to a broader, more representative range of 6.7GHz methanol maser luminosities and Galactic longitudes.

\section{Summary and conclusions}
\label{sec:Conclusion}

The distributions of outflow masses, momenta and dynamical time-scales of MMAOs have been investigated and compared to previous related studies.  We found that for an outflow mass range between $\sim 25 - 100~\rm{M_{\odot}}$, and outflow momenta between $\rm{\sim 100 - 500~M_{\odot}~km~s^{-1}}$, the MMAO distribution follows a trend similar to that found in other studies.  However, outside of these ranges, we find a comparatively smaller fraction of low-mass and momenta outflows and a higher fraction of high-mass and momenta outflows. 

The distribution of dynamical timescales shows that young MMAOs are few in number with an excess of old outflows when compared to other outflow surveys.  This indicates a dynamically older sample of outflows compared to samples not biased by their association to 6.7GHz methanol masers. Given this, it suggests that masers and outflows do not develop at the same epoch, but that masers switch on after the onset of the outflow.  This is consistent with \citet{Viti2004}'s chemical evolution models which found that large organic compounds, like methanol, are good indicators of the more evolved hot core phase, and that it is unlikely for 6.7GHz methanol masers to occur before an age of $4 \times 10^4$ yr.  

Furthermore, a modification to the first part of the evolutionary sequence for the hot core phase, presented by \citet{Codella2004}, may be required.  Without redefining the whole evolutionary cycle for high mass star formation, this study better defines the relationship of 6.7GHz methanol masers with outflows, where the masers appear at a more advanced stage of the hot core evolution than previously thought.

\section{Acknowledgements}

The James Clerk Maxwell Telescope has historically been operated by the Joint Astronomy Centre on behalf of the Science and Technology Facilities Council of the United Kingdom, the National Research Council of Canada and the Netherlands Organisation for Scientific Research. The program ID for the observations is M07AU20. We also thank the anonymous referee for constructive comments on the paper.

\label{sec:acknowledgements}

\bibliographystyle{mn2e}
\bibliography{Sources}

\end{document}